\documentclass[aps,prb,twocolumn,superscriptaddress,amsmath,amssymb]{revtex4-2}
\usepackage{graphicx}
\usepackage{dcolumn}
\usepackage{bm}
\usepackage{subfig}
\begin{document}


\title{Realizing topologically protected ghost surface polaritons by lattice transformation optics}


\author{Xianghong Kong}
\email[]{kong@nus.edu.sg}
\affiliation{Department of Electrical and Computer Engineering, National
	University of Singapore, Singapore, Singapore}
	
\author{Chuanjie Hu}
\affiliation{Shenzhen Research Institute of Xiamen University, Shenzhen 518000, China}
\affiliation{Department of Electrical and Computer Engineering, National
	University of Singapore, Singapore, Singapore}
\author{Xingsi Liu}
\affiliation{Department of Electrical and Computer Engineering, National
	University of Singapore, Singapore, Singapore}
	\author{Chunqi Zheng}
	\affiliation{Department of Electrical and Computer Engineering, National
		University of Singapore, Singapore, Singapore}
	\author{Jianfeng Chen}
\affiliation{Department of Electrical and Computer Engineering, National
	University of Singapore, Singapore, Singapore}		
\author{Huanyang Chen}
\affiliation{Shenzhen Research Institute of Xiamen University, Shenzhen 518000, China}

\author{Cheng-Wei Qiu}
\email{chengwei.qiu@nus.edu.sg}
\affiliation{Department of Electrical and Computer Engineering, National
	University of Singapore, Singapore, Singapore}

\date{\today}

\begin{abstract}
While conventional surface waves propagate along the surface and decay perpendicularly from the interface, the ghost surface polaritons show oblique propagation direction with respect to the interface. Here, we have discovered topologically protected ghost surface polaritons by applying the lattice transformation optics method to gyromagnetic photonic crystals. By introducing the transformation optics method to periodic systems, we develop the lattice transformation optics method to engineer the band structures and propagation directions of the surface polaritons. We show that a simple shear transformation on the square lattice can tailor the propagation directions with ease. The reversed ghost surface polariton is discovered by setting a negative shear factor. Interestingly, we find the topological invariant Chern number will change sign when the orientation of the Brillouin zone flipped during the transformation. Our findings open up new avenues for studying ghost surface polaritons and provide a general engineering method for periodic systems.
\end{abstract}


\maketitle

\section{Introduction}
Conventional surface polaritons propagate along the interface between two different materials and decay perpendicularly from the interface. Recently, ghost surface polaritons (GSPs) have been discovered at the interface of a bulk calcite crystal where oblique wavefronts are shown inside the calcite bulk \cite{ma2021ghost}. The name 'ghost' is in analogy with the ghost orbits in semiclassical quantization \cite{kus1993prebifurcation}. It is claimed the GSPs have longer propagation distances compared with the conventional surface polaritons. Similar results are also discovered in low symmetric monoclinic crystal \cite{hu2023real}, the interplay between periodic gratings and van der Waals crystal \cite{zhang2022unidirectionally}, deformable origami metamaterial \cite{li2022topologically}, homogeneous antiferromagnets \cite{song2022ghost}, and planar junctions between isotropic and anisotropic metasurfaces \cite{moccia2023ghost}. Despite the increasing interest in GSPs, an intuitive design to create GSPs is still waiting to be discovered.

Since the discovery of transformation optics (TO) \cite{pendry2006controlling,leonhardt2006optical}, it has become a powerful analytical tool for designing various applications such as cloak \cite{pendry2006controlling}, field concentrator \cite{rahm2008design}, optical black hole \cite{ba2022conformal}, and illusion optics \cite{liu2019illusion}. By relating the complex transformed structure to the simple original structure, an intuitive and insightful understanding of the transformed structure can be achieved. Among all the different applications designed by TO, only a very few cover the periodic structure design \cite{pendry2017compacted,galiffi2018broadband,lu2021revealing}. However, due to the existence of phase factor in the Bloch function, the band diagram of the transformed structure cannot be predicted from the original structure except at $\Gamma$ point \cite{lu2021revealing}.

In this paper, we develop the lattice transformation optics method where the TO method is applied to the periodic system to create GSPs in the gyromagnetic photonic crystal. By a simple shear transformation on the square lattice, we can change the shape of the Brillouin zone and the propagation directions of the surface polaritons. By tuning the shear factor to a negative value, the reversed GSPs are discovered where the direction of the wavefront is opposite to the direction of the energy flow. Due to the nontrivial topology of the gyromagnetic crystal \cite{wang2008reflection}, the unidirectional topologically protected GSPs can be observed at the boundary of the photonic crystal, which are immune to local disorders and defects. Furthermore, we find the Chern number would change its sign if the orientation of the Brillouin zone flips when the lattice transformation optics method is applied. The new platform we provided for engineering the band diagrams, topological invariants, and propagation directions of the surface polaritons might push forward the study of polaritons and control light in a more flexible way.

\section{\label{}Topologically protected GSPs and reversed GSPs}

\begin{figure}
	\centering
\subfloat[]{{\includegraphics[width=.45\textwidth]{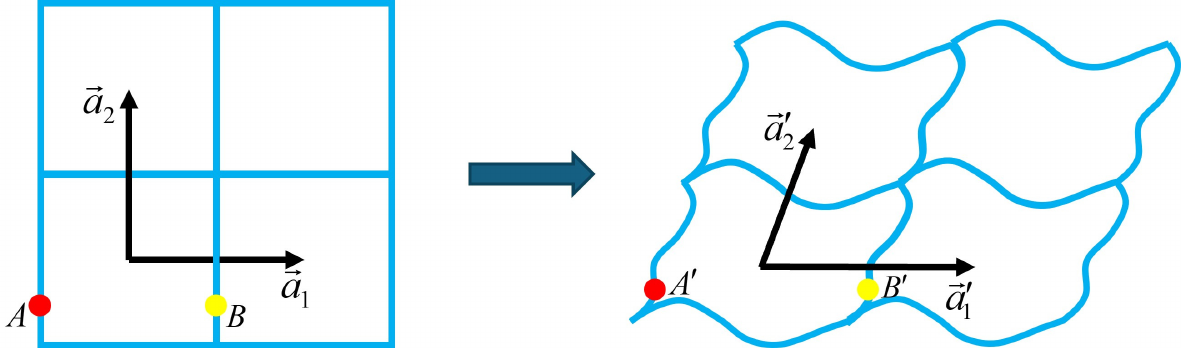}}\label{fig:fig1a}}\\

\subfloat[]{{\includegraphics[width=.54\textwidth]{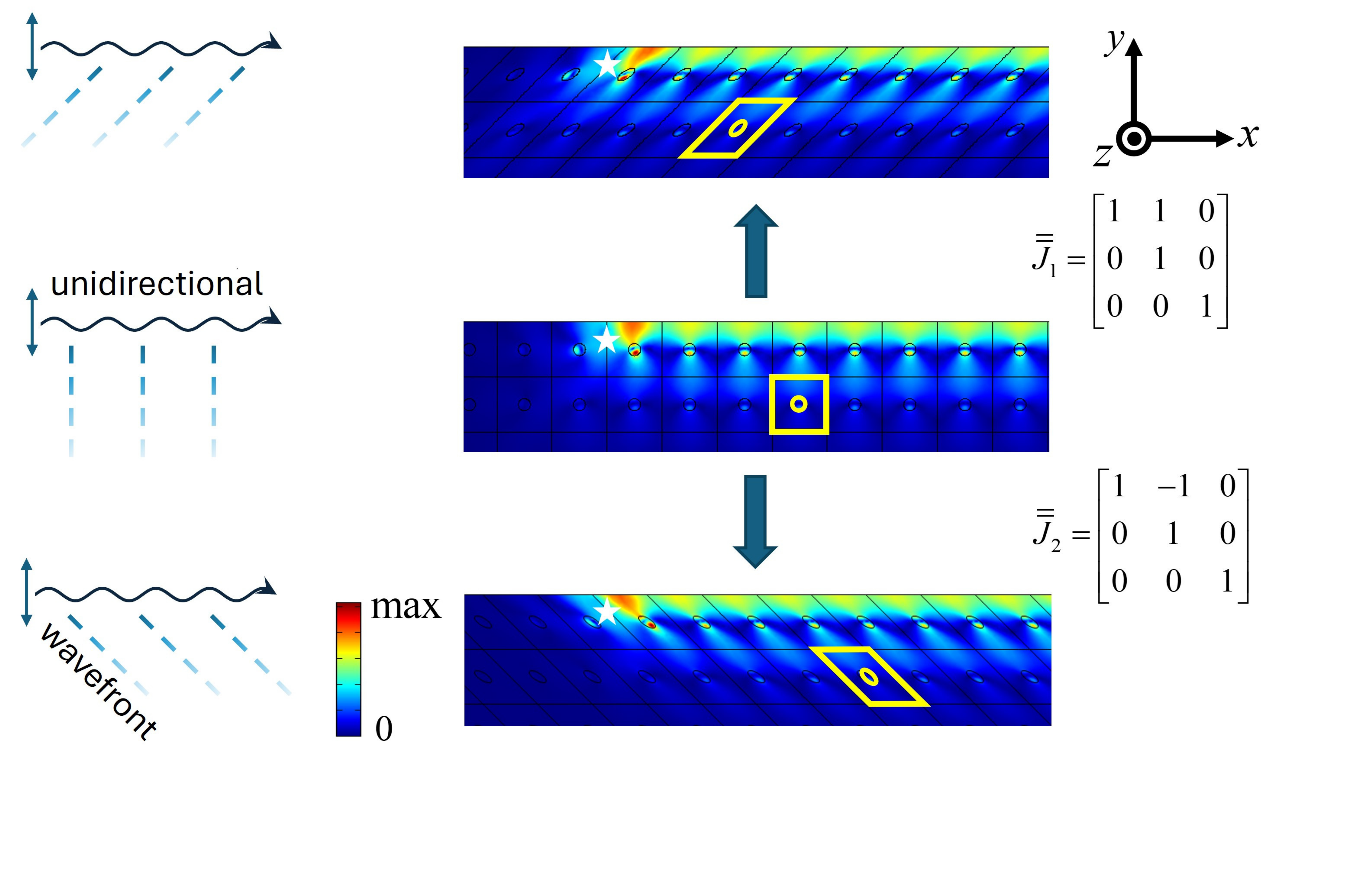}}\label{fig:fig1b}}\\

	\caption{Schematic of topologically protected GSPs and reversed GSPs designed by lattice transformation optics. (a) The square lattice is transformed into a periodic structure with curvy boundaries. (b) $|E_z|$ field distributions of the original square lattice (middle) and transformed lattice (top: transformation $\bar{\bar{J_1}}$, bottom: transformation $\bar{\bar{J_2}}$) excited by the source. The white star-shaped marks are line currents along the z-axis with normalized frequency $fa/c=0.543$. The unit cells of the original structure and transformed structures are marked by yellow. }
\end{figure}

As shown in the middle of Fig. \ref{fig:fig1b}, we consider a square lattice (period $a$) with a Yttrium-Iron-Garnet (YIG) rod in the center (for detailed material description see Appendix~\ref{appendix:materials}). The authors show in \cite{wang2008reflection} that the lowest four bands of the TM mode are well separated and nontrivial Chern numbers can be achieved in such gyromagnetic photonic crystal. Due to the bulk-boundary correspondence \cite{asboth2016short}, the unidirectional topologically protected surface mode could be discovered at the boundary of the photonic crystal, which is truncated by the perfect magnetic conductor (PMC) boundary. It is clear that the wavefront of the surface wave is perpendicular to the PMC boundary.
By applying the shear transformation \cite{jie2023shear} to the square lattices (see Appendix~\ref{appendix:materials}), we can achieve photonic crystals with parallelogram unit cells (top and bottom of Fig. \ref{fig:fig1b}). The transformation can be represented as
\begin{equation}
	\bar{\bar{J}}=\left[\begin{array}{ccc}
		1 & t_0 & 0\\
		0 & 1 & 0\\
		0 & 0 & 1
	\end{array} \right] 
	\label{eq:shear}
\end{equation}
where $t_0$ is defined as the shear coefficient. Hence, transformations $\bar{\bar{J}}_1$ and $\bar{\bar{J}}_2$ shown in Fig. \ref{fig:fig1b} can be viewed as setting $t_0=1$ and $t_0=-1$ respectively. Oblique wavefronts of the unidirectional surface waves can be observed at the interface, which proves the existence of the topologically protected GSPs. Interestingly, similar to the reversed Cherenkov radiation \cite{hu2020directing,guo2023mid}, the reversed GSPs can be realized easily by setting negative shear coefficients $t_0$ in the shear transformation. The direction of the radiation energy flow ($-x$) is opposite to the wave vector ($+x$), which can be verified in Fig. \ref{fig:fig1b}.

\section{Lattice transformation optics}

\begin{figure}
	\centering
	\subfloat[]{{\includegraphics[width=.54\textwidth]{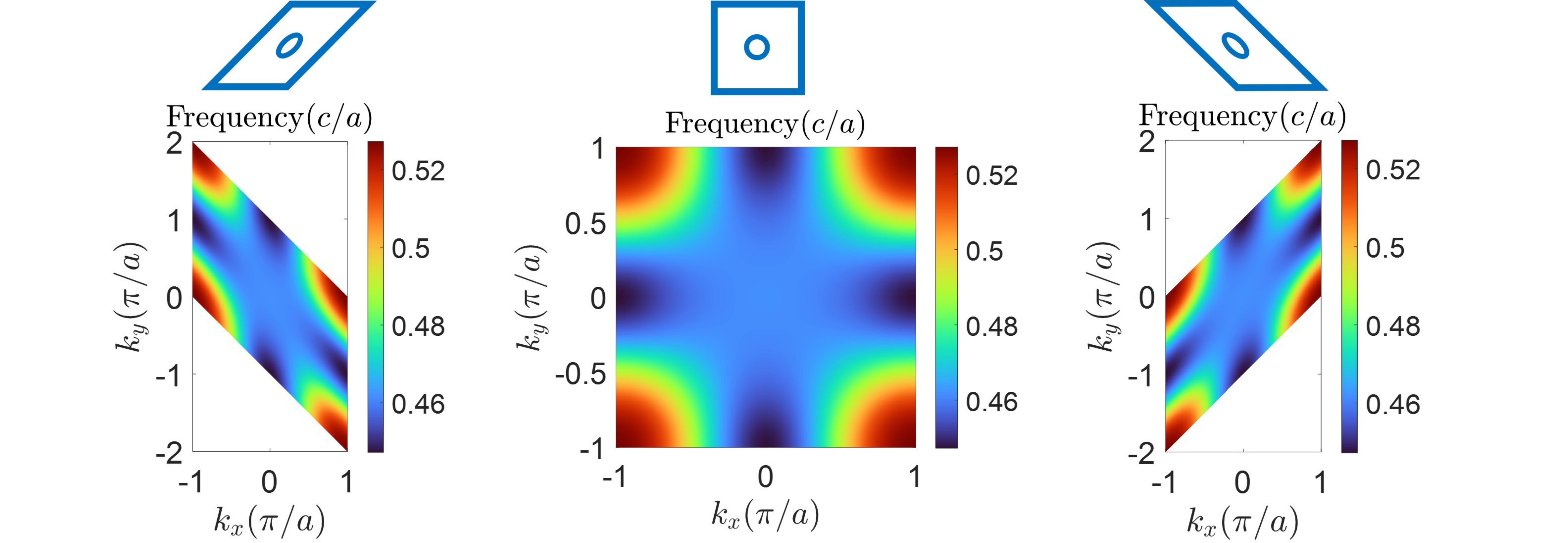}}\label{fig:fig2a}}\\
	
	\subfloat[]{{\includegraphics[width=.28\textwidth]{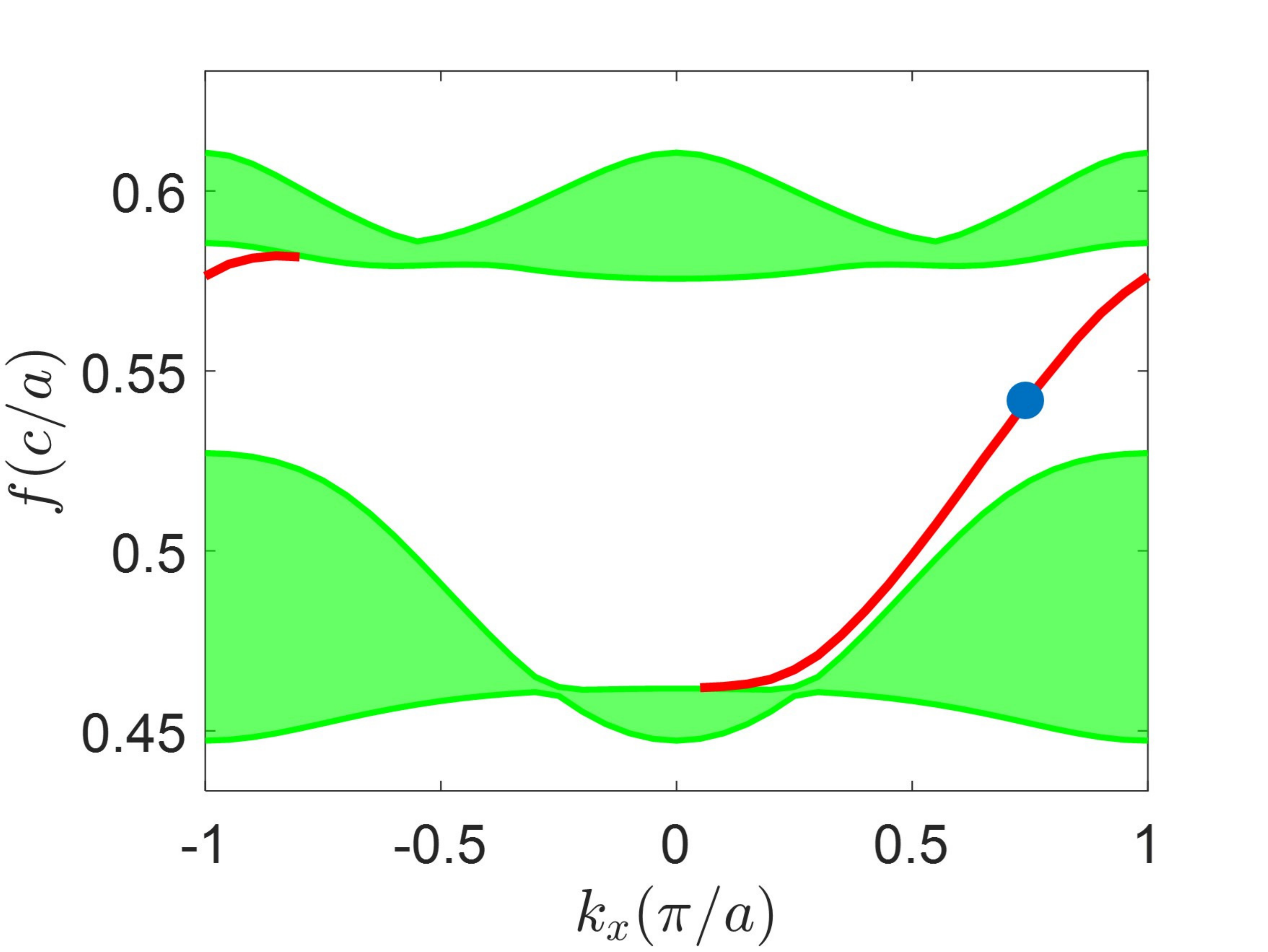}}\label{fig:fig2b}}
	\subfloat[]{{\includegraphics[width=.24\textwidth]{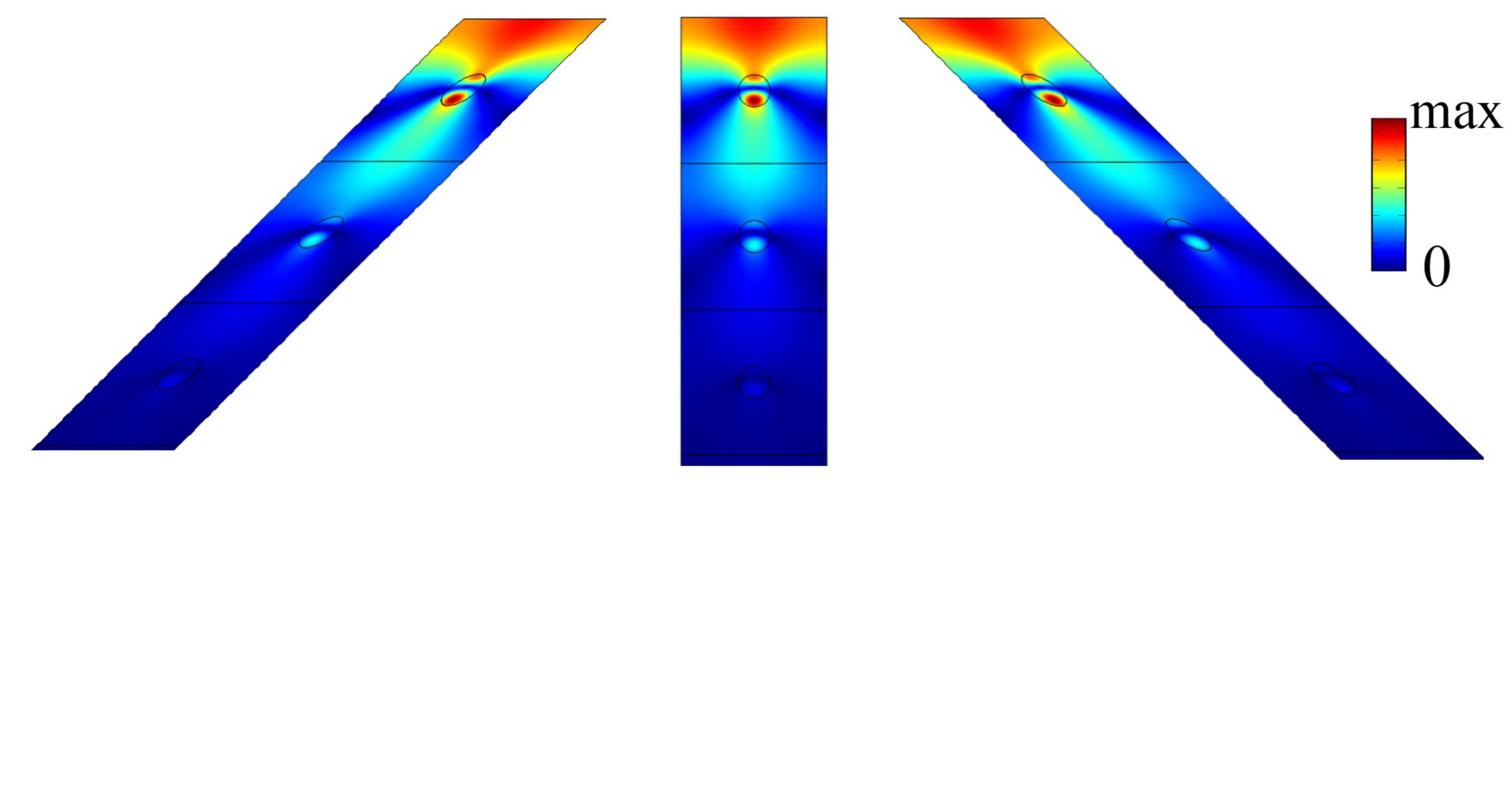}}\label{fig:fig2c}}\\

	\caption{Band analysis of the original unit cell and the transformed unit cells. (a) Band diagrams of the second band of the corresponding unit cells shown in blue. Middle: A square lattice  with a YIG rod in the center. Left and right: The diamond lattice is transformed from the square lattice in the middle under the matrix $\bar{\bar{J_1}}$ and $\bar{\bar{J_2}}$ respectively. (b) Projected band diagram of the second and third band of the three unit cells shown in (a). The red curve is the dispersion relation of the super-cells formed by the three unit cells. The blue dot is located at $k_x=0.75\pi/a, fa/c=0.543$. (c) $|E_z|$ field distributions of surface modes at the blue dot in (b)}
	
\end{figure}

The Bloch theorem shows the electric field should have the form $E_z=e^{i\vec{k}^T\cdot\vec{r}}u_{\vec{k}}\left(\vec{r} \right) $ where $u_{\vec{k}}\left(\vec{r}\right)$ is a periodic function and satisfies $u_{\vec{k}}\left(\vec{r} \right)=u_{\vec{k}}\left(\vec{r}+\vec{a}_1 \right)=u_{\vec{k}}\left(\vec{r}+\vec{a}_2 \right)$. Here, $\vec{a}_1=(1,0)^T$ and $\vec{a}_2=(0,1)^T$ are lattice vectors of the square crystal. The phase difference between point $A$ and point $B$ (see Fig. \ref{fig:fig1a}) in the original space should be the same compared with the points $A^{\prime}$ and $B^{\prime}$ in the transformed space. Hence, we can conclude that:
\begin{subequations}
	\begin{eqnarray}
		e^{i\vec{k}^{\prime T}\cdot\vec{a}^{\prime}_1}=e^{i\vec{k}^{ T}\cdot\vec{a}_1}
	\end{eqnarray}
	\begin{eqnarray}
		e^{i\vec{k}^{\prime T}\cdot\vec{a}^{\prime}_2}=e^{i\vec{k}^{ T}\cdot\vec{a}_2}
	\end{eqnarray}
	\label{eq:relation_k}
\end{subequations}
Although Eq.~(\ref{eq:relation_k}) is not only valid in the linear transformation, we want to emphasize that not all the nonlinear transformations can match the condition shown in Eq.~(\ref{eq:relation_k}). The existence of the lattice vector $\vec{a}^{\prime}_1$ and $\vec{a}^{\prime}_2$ in the transformed space indicates two pairs of the periodic boundary should be reserved during the transformation. The line shape of the periodic boundary does not have to be straight (see Appendix~\ref{appendix:curved}).
In our linear case, the relation between $\vec{a}_1$, $\vec{a}_2$ and $\vec{a}_1^{\prime}$, $\vec{a}_2^{\prime}$ can be expressed as:
\begin{equation}
	\left[\vec{a}_1^{\prime}, \vec{a}_2^{\prime}\right]=\begin{bmatrix}
		\dfrac{\partial x^{\prime}}{\partial x}&\dfrac{\partial x^{\prime}}{\partial y}\\
		\dfrac{\partial y^{\prime}}{\partial x}&\dfrac{\partial y^{\prime}}{\partial y}
	\end{bmatrix}\left[ \vec{a}_1, \vec{a}_2\right] 
	\label{eq:linear_transform}	
\end{equation}
Combining Eq.~(\ref{eq:relation_k}) and Eq.~(\ref{eq:linear_transform}), we can figure out how the k space transforms:
\begin{equation}
	\vec{k}=\begin{bmatrix}
		\dfrac{\partial x^{\prime}}{\partial x}&\dfrac{\partial y^{\prime}}{\partial x}\\
		\dfrac{\partial x^{\prime}}{\partial y}&\dfrac{\partial y^{\prime}}{\partial y}
	\end{bmatrix}\vec{k}^{\prime}
	\label{eq:transform_k}
\end{equation}
As shown in Fig. \ref{fig:fig2a}, the four corners of the Brillouin zone of the original square lattice are $(-\pi/a, -\pi/a)$, $(\pi/a, -\pi/a)$, $(\pi/a, \pi/a)$, and $(-\pi/a, \pi/a)$. When the transformation matrix is $\bar{\bar{J}}_1=\begin{bmatrix}
	1&1&0\\
	0&1&0\\
	0&0&1
\end{bmatrix}$, we can figure out the four corners of the transformed Brillouin zone $(-\pi/a, 0)$, $(\pi/a, -2\pi/a)$, $(\pi/a, 0)$, and $(-\pi/a, 2\pi/a)$ by applying Eq.~(\ref{eq:transform_k}). Similarly, for the transformation matrix $\bar{\bar{J}}_2=\begin{bmatrix}
	1&-1&0\\
	0&1&0\\
	0&0&1
\end{bmatrix}$, we can find out the corners of the transformed Brillouin zone are $(-\pi/a, -2\pi/a)$, $(\pi/a, 0)$, $(\pi/a, 2\pi/a)$, and $(-\pi/a, 0)$. The relation between $(k_x^{\prime},k_y^{\prime})$ and $(k_x,k_y)$ for the shear transformation shown in Eq.~(\ref{eq:shear}) can be represented as $\begin{bmatrix}
	k_x^{\prime}\\k_y^{\prime}
\end{bmatrix}=\begin{bmatrix}
	1&0\\
	-t_0&1
\end{bmatrix}\begin{bmatrix}
	k_x\\k_y
\end{bmatrix}$. No matter how complicated the transformation may behave in the real space, the reaction in the k space is always a simple linear transformation as shown in Eq.~(\ref{eq:relation_k}) since it is only related to the transformation of the lattice vectors. As shown in Fig. \ref{fig:fig2b}, both the original structure and the transformed structures share the same projected band diagram. It is due to the fact that $k_x^{\prime}=k_x$ during the shear transformations. The dispersion relations of the surface modes for all three super-cells are also coincident since they only relate to $k_x$, which is invariant during the transformations. The $|E_z|$ field distributions of the surface modes at $k_x=0.75\pi/a$, $fa/c=0.543$ are plotted in Fig. \ref{fig:fig2c}. Comparing Fig. \ref{fig:fig2c} with Fig. \ref{fig:fig1b} we can conclude that the GSPs excited by the line currents are exactly the eigenmodes of the super-cells since they have same field distributions and normalized frequencies. By assuming the k vector in the original structure as $\begin{bmatrix}
	k_x\\k_y
\end{bmatrix}=\begin{bmatrix}
	0.75\pi/a\\-i|\Im(k_y)|
\end{bmatrix}$, the $k^{\prime}$ vector in the transformed structure can be calculated as $\begin{bmatrix}
	k_x^{\prime}\\k_y^{\prime}
\end{bmatrix}=\begin{bmatrix}
	0.75\pi/a\\-0.75t_0\pi/a-i|\Im(k_y)|
\end{bmatrix}$, which matches the wavefronts for different shear coefficients and can be checked by observing Fig. \ref{fig:fig2c} easily. The group velocity $v_g$ can be defined as $v_g=\partial\omega/\partial k$, which is always positive in x direction judging from the slope of the red curve in Fig. \ref{fig:fig2b}. This explains the unidirectional propagation of the GSPs and will be further discussed in the following section.

\section{Topological analysis}

\begin{figure}
	\centering
	\includegraphics[width=0.53\textwidth]{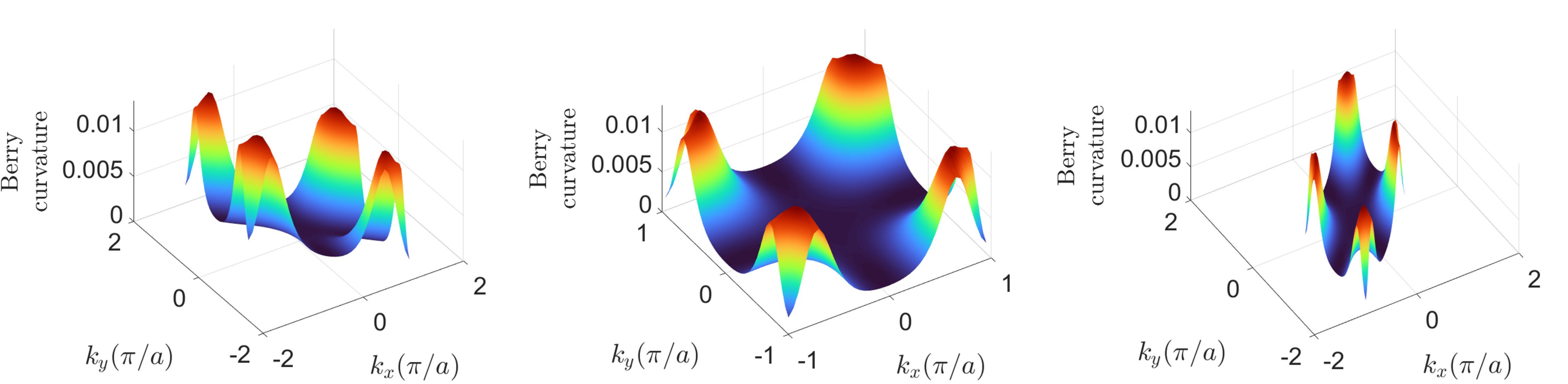}
	\caption{Berry curvatures of the second band of the original structure and the transformed structures. Middle: Original square lattice. Left and right: The transformed lattices under shear transformations $\bar{\bar{J_1}}$, $\bar{\bar{J_2}}$ respectively.}
	\label{fig:fig3}
\end{figure}
It is well known that the gyromagnetic material will show nontrivial Chern numbers for different bands \cite{wang2008reflection} since it breaks the time-reversal symmetry. The topological property ensures the existence and unidirectional propagation of the GSPs. Due to its importance, it would be interesting to investigate how the Chern number would change when the lattice transformation optics is applied.

Through detailed derivation (see Appendix~\ref{appendix:chern}), we discover the relation between the Chern number in the transformed space and the original space:
\begin{equation}
	C^{\prime}\!=\mathrm{sign}\left(\mathrm{det}\left( \dfrac{\partial\left(k_x^{\prime},k_y^{\prime} \right) }{\partial\left(k_x,k_y \right) }\right)  \right) C
	\label{eq:transform_chern_manu}
\end{equation}
According to Eq.~(\ref{eq:transform_chern_manu}), we can conclude that the Chern number in the transformed space will change its sign compared with the original space when the orientation of the Brillouin zone is flipped after the transformation. For our shear transformation, the $\mathrm{det}\left( \dfrac{\partial\left(k_x^{\prime},k_y^{\prime} \right) }{\partial\left(k_x,k_y \right) }\right)=1$ whatever the value of $t_0$ is (see Appendix~\ref{appendix:example} for the sign-changing example). Hence, the Chern number remains the same and the direction of the surface wave is also invariant after the transformation as shown in Fig. \ref{fig:fig1b}. 

As shown in Fig. \ref{fig:fig3}, the Berry curvatures of the original structure and the transformed structures are plotted. By summing up the Berry curvature in the whole Brillouin zone, we can achieve the Chern number, which is $C=1$ for all three cases. 

\section{Tuning the shear coefficient}

\begin{figure}
	\centering
	\hspace*{-0.7cm}
	\subfloat[]{{\includegraphics[width=.27\textwidth]{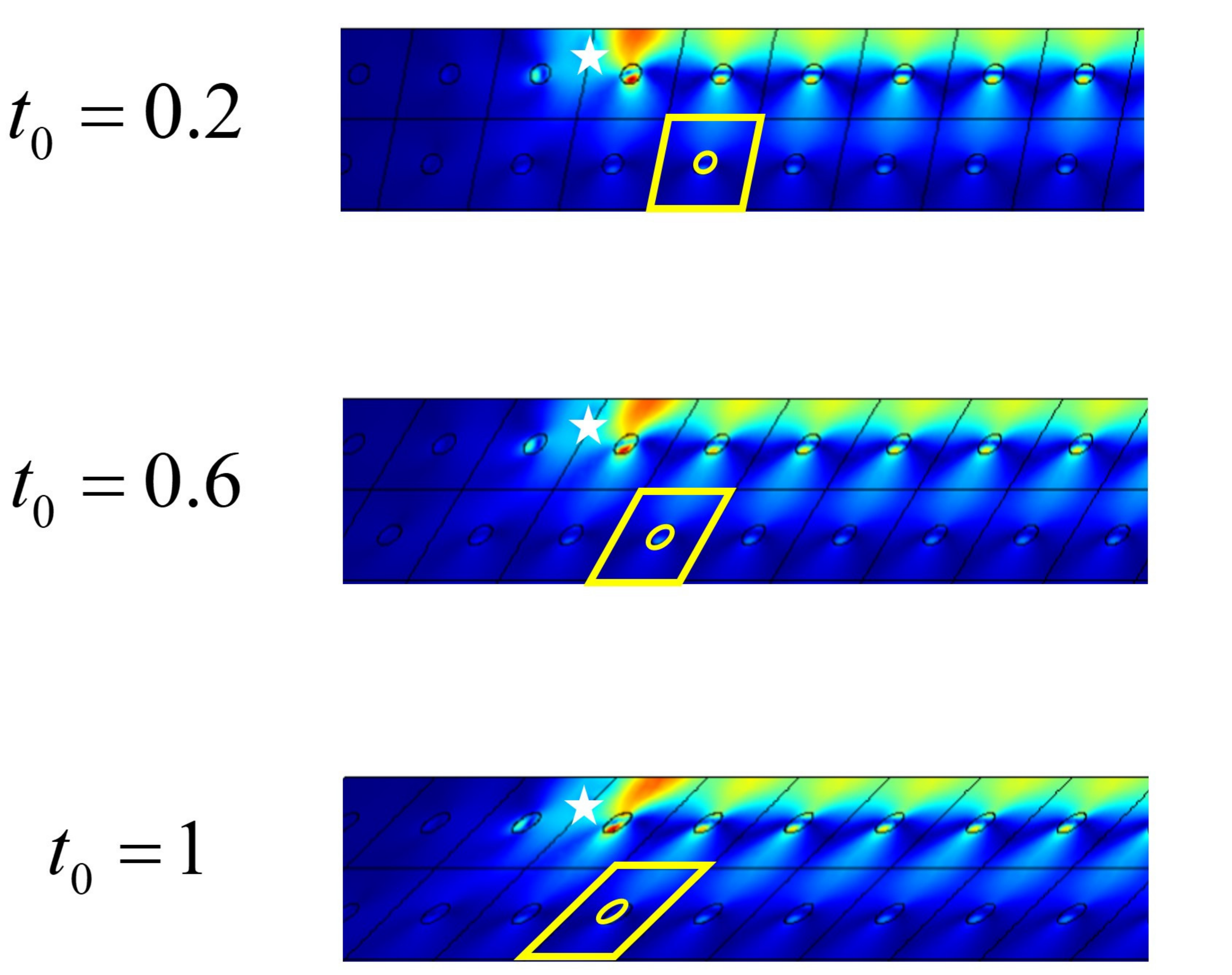}}\label{fig:fig4a}}
	\hspace*{-0.4cm}
	\subfloat[]{{\includegraphics[width=.28\textwidth]{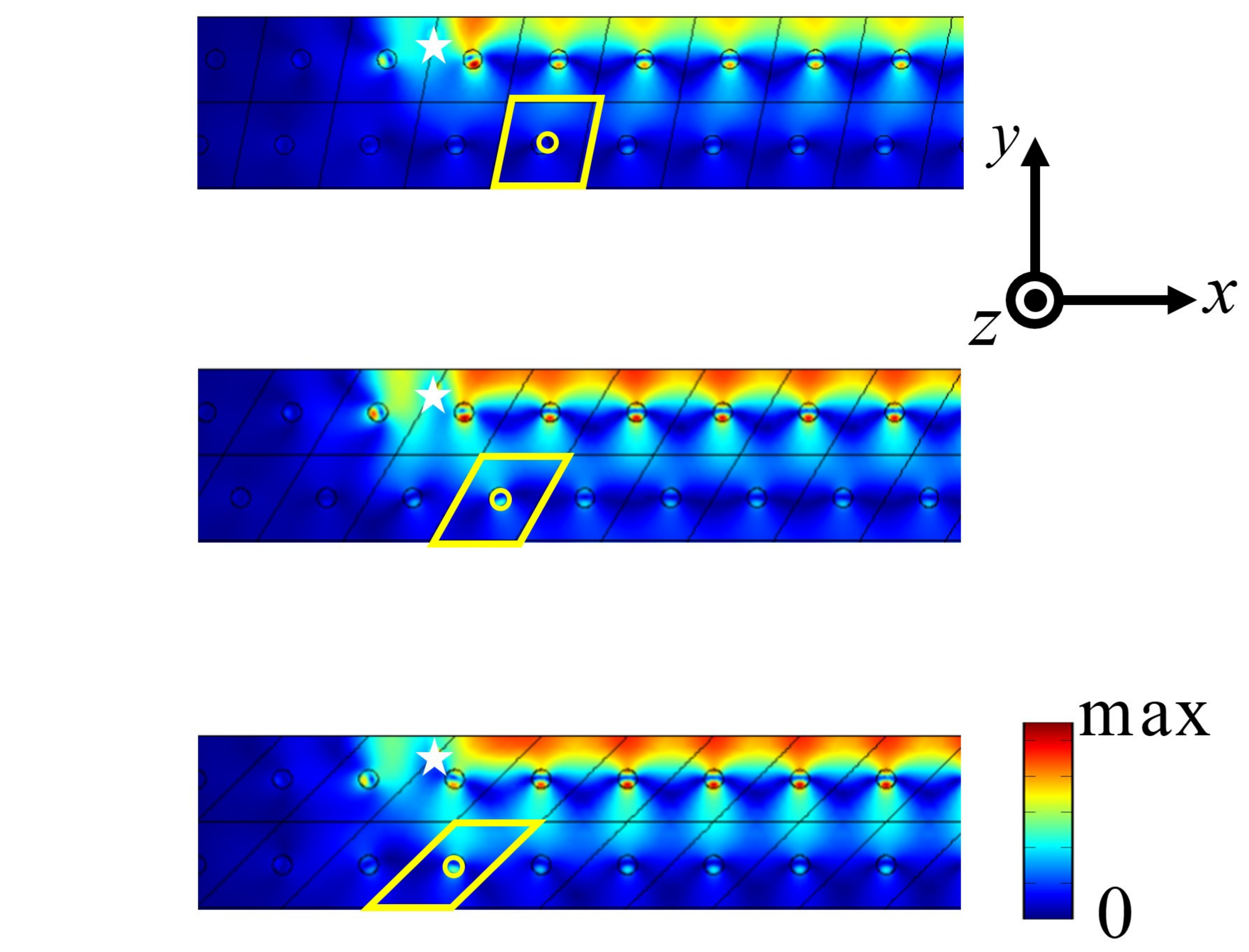}}\label{fig:fig4b}}\\
	\caption{A comparison between the GSPs and conventional surface polaritons when tuning the shear coefficient. (a) The GSPs created by the lattice transformation optics, where both the shape of the YIG rod and the material parameters are transformed according to the TO. (b) Only the shape of lattice is transformed, while the shape of the rod and the material parameters remain the same as the original structure. The white star-shaped marks are line currents along the z-axis with normalized frequency $fa/c=0.543$. The unit cells of the transformed structures are marked by yellow. }
	\label{fig:fig4} 
\end{figure}

In Fig. \ref{fig:fig4}, we make a comparison between the GSPs created by the lattice transformation optics and the conventional surface polaritons for different shear coefficients $t_0$. The normalized frequency of the point source is $fa/c=0.543$. As shown in Fig. \ref{fig:fig4a}, the wavefront can be tuned by the shear coefficient and the quantitative relation between $k^{\prime}$ vector and $k$ vector is $\begin{bmatrix}
	k_x^{\prime}\\k_y^{\prime}
\end{bmatrix}=\begin{bmatrix}
	k_x\\-t_0k_x-i|\Im(k_y)|
\end{bmatrix}$ by applying Eq.~(\ref{eq:transform_k}). However, a simple transformation of lattice without changing the material parameters according to the TO will not change the direction of the wavefront, which can be verified in Fig. \ref{fig:fig4b}. The comparison highlights the importance of the lattice transformation optics method when engineering the topologically protected GSPs.

\section{Discussion and outlook}

Here, we show the realization of topologically protected GSPs and reversed GSPs by applying the lattice transformation optics method. Compared with previous GSPs schemes, our method shows an ultrawide tuning range of wavefront, which may open new avenues for controllable surface polaritons transfer, sensing, and energy transport. Moreover, the lattice transformation optics we developed can be applied to tune the band diagrams, eigenfield distributions, and topological invariants of the periodic structures, which may be a potential tool for flat band design \cite{nguyen2022magic, tang2020photonic} and 3D Chern vector design \cite{devescovi2022vectorial, liu2022topological}.

\appendix
\section{\label{appendix:materials}Materials and methods}

\subsection{The photonic crystals description}
We consider a square lattice (period $a$) with a YIG rod ($r=0.11a$) in the center. The permittivity of the YIG rod is $\epsilon=15\epsilon_0$ and the permeability tensor is \cite{wang2008reflection}:
\begin{equation}
	\bar{\bar{\mu}}	=\left[ \begin{array}{ccc}
		\mu & i\kappa & 0 \\
		-i\kappa & \mu & 0\\
		0 & 0 & \mu_0
	\end{array}\right]\nonumber
\end{equation}
where $\mu=14\mu_0$ and $\kappa=12.4\mu_0$. After applying the shear transformation, the permittivity and permeability of the YIG rod are transformed into \cite{pendry2006controlling}
\begin{equation}
	\bar{\bar{\epsilon}}^{\prime}=\dfrac{\bar{\bar{J}}\cdot\bar{\bar{\epsilon}}\cdot\bar{\bar{J}}^T}{\mathrm{det}\left(\bar{\bar{J}} \right)}=\epsilon\left[ \begin{array}{ccc}
		1+t_0^2 & t_0 & 0\\
		t_0 & 1 & 0\\
		0 & 0 & 1
	\end{array}\right] 
	\nonumber
\end{equation}
\begin{equation}
	\bar{\bar{\mu}}^{\prime}=\dfrac{\bar{\bar{J}}\cdot\bar{\bar{\mu}}\cdot\bar{\bar{J}}^T}{\mathrm{det}\left(\bar{\bar{J}} \right) }=\left[\begin{array}{ccc}
		(1+t_0^2)\mu & t_0\mu+i\kappa & 0\\
		t_0\mu-i\kappa & \mu & 0\\
		0 & 0 & \mu_0
	\end{array} \right] 
	\nonumber
\end{equation}
Similarly, the vacuum surrounding the rod in the original square lattice will also transformed into 
$\epsilon_0\left[ \begin{array}{ccc}
	1+t_0^2 & t_0 & 0\\
	t_0 & 1 & 0\\
	0 & 0 & 1
\end{array}\right] $ and $\mu_0\left[ \begin{array}{ccc}
	1+t_0^2 & t_0 & 0\\
	t_0 & 1 & 0\\
	0 & 0 & 1
\end{array}\right] $. The geometry of the circle $x^2+y^2=r^2$ becomes an ellipse $(x^{\prime}-t_0y^{\prime})^2+y^{\prime 2}=r^2$.

\subsection{Numerical simulations}
All the simulations in the paper are performed by the COMSOL Multiphysics. For the driven mode simulations, we apply the PMC boundary on the top and scattering boundary condition (SBC) for the other three boundaries, such as Fig. \ref{fig:fig1b}, Fig. \ref{fig:fig4}. For the eigenmode simulations of the super-cells, we set PMC on the top and SBC on the bottom and a pair of periodic boundary for the left and right boundaries, such as Fig. \ref{fig:fig2c}.

\section{\label{appendix:curved}Curved periodic boundary }

\begin{figure}
	\centering
	\includegraphics[width=.5\textwidth]{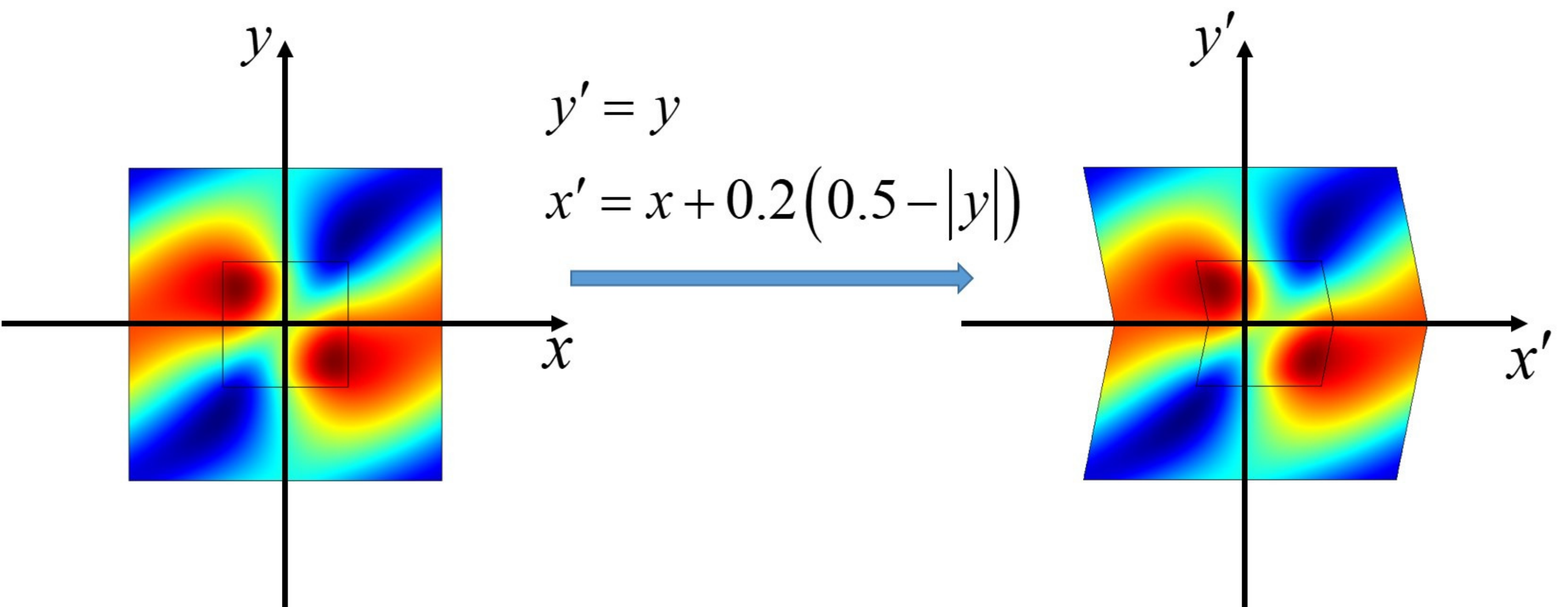}
	
	\caption{Transformation from a square lattice to a lattice with curved periodic boundaries. The period of the square lattice is $a=1\mathrm{m}$ and the side length is $0.4\mathrm{m}$ for the smaller square. The material is set as $\epsilon=2\epsilon_0$, $\mu=2\mu_0$ for the smaller square and it is surrounded by the vacuum. $|E_z|$ field distributions at $k_x=\pi/3$, $k_y=\pi/2$, and $f=0.74c/a$ are plotted for the original structure and transformed structure.}
	\label{fig:figs1}
\end{figure}

As shown in Fig. \ref{fig:figs1}, a square lattice can be transformed into a lattice with curved periodic boundaries. The period of the square is $1\mathrm{m}$ and the side length of the smaller square is $0.4\mathrm{m}$. The permittivity and permeability of the smaller square are $\epsilon=2\epsilon_0$ and $\mu=2\mu_0$ respectively. The transformation matrix is $\bar{\bar{J}}=\begin{bmatrix}
	1&\mp0.2&0\\
	0&1&0\\
	0&0&1
\end{bmatrix}$ where '$-$' is for the domain $y>0$ and '$+$' is for the domain $y<0$. The corresponding material transformation follows the rules governed by \cite{pendry2006controlling} 
\begin{equation}
	\bar{\bar{\epsilon}}^{\prime}=\dfrac{\bar{\bar{J}}\cdot\bar{\bar{\epsilon}}\cdot\bar{\bar{J}}^T}{\mathrm{det}\left(\bar{\bar{J}} \right) },\,\,\,\,\bar{\bar{\mu}}^{\prime}=\dfrac{\bar{\bar{J}}\cdot\bar{\bar{\mu}}\cdot\bar{\bar{J}}^T}{\mathrm{det}\left(\bar{\bar{J}} \right) }
	\label{eq:transform_material}
\end{equation}
Since the transformed lattice vectors are the same as the original lattice vectors ($\vec{a}_1^{\prime}=\vec{a}_1$, $\vec{a}_2^{\prime}=\vec{a}_2$), the k space also keeps the same $\vec{k}^{\prime}=\vec{k}$ according to the Eq.~(\ref{eq:relation_k}). The $|E_z|$ field distribution of the original structure is plotted at $k_x=\pi/3$, $k_y=\pi/2$, and $f=0.74c/a$ in Fig. \ref{fig:figs1}. It matches to $k_x^{\prime}=\pi/3$, $k_y^{\prime}=\pi/2$, and $f^{\prime}=0.74c/a$ in the transformed space. By comparing the field distributions, we find they exactly follow the rules given in 	
\begin{equation}
	\vec{E}^{\prime}=\left( \bar{\bar{J}}^T\right)^{-1}\cdot\vec{E},\,\,\,\,\vec{H}^{\prime}=\left( \bar{\bar{J}}^T\right)^{-1}\cdot\vec{H}
	\label{eq:transform_field}
\end{equation} 
For more complicated transformed periodic boundaries, we can use the piecewise linear boundaries to approximate them and run Comsol simulations to help understand the eigenfields.

\section{\label{appendix:chern}Chern number calculation under lattice transformation optics}

Here, we will derive the Chern number of the original space and transformed space. 
The Berry connection in the 2D photonic crystal system can be written as \cite{wang2008reflection}:
\begin{subequations}
	\begin{eqnarray}
		A_x=\iint E_i^*\epsilon^{ij}\partial_{k_x}E_j\,\mathrm{d} x \, \mathrm{d} y
		\label{eq:berry1}
	\end{eqnarray}
	\begin{eqnarray}
		A_y=\iint E_i^*\epsilon^{ij}\partial_{k_y}E_j\,\mathrm{d} x \, \mathrm{d} y
		\label{eq:berry2}
	\end{eqnarray}
	\label{eq:berry12}
\end{subequations}
where $\epsilon^{ij}$, $E_i$ represent the 3 by 3 permittivity and 3 by 1 electric field in the tensor form. In our 2D case, we have $E_1=0, E_2=0, E_3=E_z$. Repeated indices $i$ and $j$ are summed over according to the Einstein summation rules. Hence, the Chern number in the original space is:
\begin{eqnarray}
	C & = & \frac{1}{2\pi i}\iint\left(\partial_{k_x}A_y-\partial_{k_y}A_x \right)\mathrm{d} k_x \, \mathrm{d} k_y \nonumber\\
	& = & \frac{1}{2\pi i}\iiiint(\partial_{k_x}E_i^*\epsilon^{ij}\partial_{k_y}E_j\nonumber\\& -& \partial_{k_y}E_i^*\epsilon^{ij}\partial_{k_x}E_j )
	\mathrm{d} x \, \mathrm{d} y \, \mathrm{d} k_x \, \mathrm{d} k_y  
	\label{eq:chern_origin}
\end{eqnarray}
We assume the electric field is normalized before transformation, which means:

\begin{equation}
	\iint E_i^*\epsilon^{ij} E_j\mathrm{d}x \,\mathrm{d}y=1
	\label{eq:normalized_original}
\end{equation}
After the transformation, the norm of the above expression changes into

\begin{eqnarray}
	\iint E_{i^\prime}^*\epsilon^{i^{\prime}j^{\prime}}E_{j^{\prime}}\mathrm{d}x^{\prime}\mathrm{d}y^{\prime} & = &\iint  E_i^*J_{i^{\prime}}^i\nonumber\\
	\dfrac{J_p^{i^{\prime}}\epsilon^{pq}J_q^{j^{\prime}}}{\mathrm{det}\left(\dfrac{\partial\left( x^{\prime},y^{\prime},z^{\prime}\right) }{\partial\left( x,y,z\right) } \right) } J_{j^{\prime}}^j &E_j &\left| \mathrm{det}\left(\dfrac{\partial\left(x^{\prime},y^{\prime} \right) }{\partial\left( x,y\right) } \right)  \right|  \mathrm{d}x\mathrm{d}y \nonumber\\
	& = & \dfrac{\left| \mathrm{det}\left(\dfrac{\partial\left(x^{\prime},y^{\prime} \right) }{\partial\left( x,y\right) } \right)  \right|}{\mathrm{det}\left(\dfrac{\partial\left( x^{\prime},y^{\prime},z^{\prime}\right) }{\partial\left( x,y,z\right) } \right)}
	\label{eq:normalized_transform}
\end{eqnarray}
The summation of two Jacobian matrices can be simplified as $J^i_{i^{\prime}}J^{i^{\prime}}_p=\delta_p^i$. Also, let's assume the transformation is linear, which means the Jacobian matrix is independent of the integration variable $x$, $y$. Then we can immediately get Eq.~(\ref{eq:normalized_transform}) by substituting Eq.~(\ref{eq:normalized_original}) into it. However, we want to emphasize that even for some special nonlinear transformation cases, Eq.~(\ref{eq:normalized_transform}) can still be valid since in 2D photonic crystal the coupling between coordinates $x,y$ and $z$ is neglected. The determinant of the Jacobian matrix can be decomposed into 
\begin{equation}
	\mathrm{det}\left( \dfrac{\partial\left(  x^{\prime},y^{\prime},z^{\prime}\right) }{\partial\left( x,y,z\right) }\right) =	\mathrm{det}\left( \dfrac{\partial\left(  x^{\prime},y^{\prime}\right) }{\partial\left( x,y\right) }\right)\dfrac{\mathrm{d}z^{\prime}}{\mathrm{d}z}\nonumber
\end{equation}
As long as the sign of $\mathrm{det}\left( \dfrac{\partial\left(  x^{\prime},y^{\prime}\right) }{\partial\left( x,y\right) }\right)$ does not change in the whole integration area, we can still take out the term $\dfrac{\left| \mathrm{det}\left(\dfrac{\partial\left(x^{\prime},y^{\prime} \right) }{\partial\left( x,y\right) } \right)  \right|}{\mathrm{det}\left(\dfrac{\partial\left( x^{\prime},y^{\prime},z^{\prime}\right) }{\partial\left( x,y,z\right) } \right)}$ and get Eq.~(\ref{eq:normalized_transform}).
Hence, the normalized electric field after transformation can be expressed as
\begin{equation}
	\iint \dfrac{\mathrm{det}\left(\dfrac{\partial\left( x^{\prime},y^{\prime},z^{\prime}\right) }{\partial\left( x,y,z\right) } \right)}{\left| \mathrm{det}\left(\dfrac{\partial\left(x^{\prime},y^{\prime} \right) }{\partial\left( x,y\right) } \right)  \right|}E_{i^\prime}^*\epsilon^{i^{\prime}j^{\prime}}E_{j^{\prime}}\mathrm{d}x^{\prime}\mathrm{d}y^{\prime}=1
\end{equation}
Similar to Eq.~(\ref{eq:berry12}), the Berry connection defined in the transformed space is
\begin{eqnarray}
	A_x^{\prime}\!\!=\!\!	\iint \dfrac{\mathrm{det}\left(\dfrac{\partial\left( x^{\prime},y^{\prime},z^{\prime}\right) }{\partial\left( x,y,z\right) } \right)}{\left| \mathrm{det}\left(\dfrac{\partial\left(x^{\prime},y^{\prime} \right) }{\partial\left( x,y\right) } \right)  \right|}E_{i^\prime}^*\epsilon^{i^{\prime}j^{\prime}}\partial _{k_x^{\prime}}E_{j^{\prime}}\mathrm{d}x^{\prime}\mathrm{d}y^{\prime}
	\label{eq:berry1'_1}
\end{eqnarray}
Here the normalization term $\dfrac{\mathrm{det}\left(\dfrac{\partial\left( x^{\prime},y^{\prime},z^{\prime}\right) }{\partial\left( x,y,z\right) } \right)}{\left| \mathrm{det}\left(\dfrac{\partial\left(x^{\prime},y^{\prime} \right) }{\partial\left( x,y\right) } \right)  \right|}$ does not relate to variable $k_x^{\prime}$, which means it can be taken out from the derivative with respect to $k_x^{\prime}$ and put at the front of the integral term as shown in Eq.~(\ref{eq:berry1'_1}). Following the same procedure as Eq.~(\ref{eq:normalized_transform}), let's replace the electric field $E_j^{\prime}$ and permittivity $\epsilon^{i^{\prime}j^{\prime}}$ in the transformed space with the electric field $E_j$ and permittivity $\epsilon^{ij}$ in the original space according to the transformation optics, we can easily get:
\begin{eqnarray}
	A_x^{\prime}=\iint E_i^*\epsilon^{ij}\partial_{k_x^{\prime}}E_j\,\mathrm{d} x \, \mathrm{d} y
	\label{eq:berry1'}
\end{eqnarray}
Similarly,
\begin{eqnarray}
	A_y^{\prime}=\iint E_i^*\epsilon^{ij}\partial_{k_y^{\prime}}E_j\,\mathrm{d} x \, \mathrm{d} y
	\label{eq:berry2'}
\end{eqnarray}
The Chern number after transformation can be calculated as

\begin{widetext}
	\begin{eqnarray}
		\!\!\!\!\!\!	C^{\prime}\!& = & \!\frac{1}{2\pi i}\iint\left(\partial_{k_x^{\prime}}A_y^{\prime}-\partial_{k_y^{\prime}}A_x^{\prime} \right)\mathrm{d} k_x^{\prime} \mathrm{d} k_y^{\prime}\! =\!\frac{1}{2\pi i}\iiiint \!\left( \partial_{k_x^{\prime}}E_i^*\epsilon^{ij}\partial_{k_y^{\prime}}E_j \!-\!\partial_{k_y^{\prime}}E_i^*\epsilon^{ij}\partial_{k_x^{\prime}}E_j\right)\! \left| \dfrac{\partial\left(k_x^{\prime},k_y^{\prime} \right) }{\partial\left(k_x,k_y \right) }\right| \mathrm{d}x\mathrm{d}y\mathrm{d}k_x\mathrm{d}k_y\nonumber\\
		& = & \!\frac{1}{2\pi i}\iiiint \!\dfrac{\left( \partial_{k_x}E_i^*\epsilon^{ij}\partial_{k_y}E_j \!-\!\partial_{k_y}E_i^*\epsilon^{ij}\partial_{k_x}E_j\right)}{\mathrm{det}\left( \dfrac{\partial\left(k_x^{\prime},k_y^{\prime} \right) }{\partial\left(k_x,k_y \right) }\right) }\! \left| \dfrac{\partial\left(k_x^{\prime},k_y^{\prime} \right) }{\partial\left(k_x,k_y \right) }\right| \mathrm{d}x\mathrm{d}y\mathrm{d}k_x\mathrm{d}k_y=\mathrm{sign}\left(\mathrm{det}\left( \dfrac{\partial\left(k_x^{\prime},k_y^{\prime} \right) }{\partial\left(k_x,k_y \right) }\right)  \right) C
		\label{eq:chern_transformed}
	\end{eqnarray}
\end{widetext}
As shown in Eq.~(\ref{eq:chern_transformed}), the Chern number can change its sign after transformation according to the change of the orientation of the Brillouin zone.

\section{\label{appendix:example}Example of sign changing of the Chern number under lattice transformation optics}

\begin{figure}
	\centering
	\subfloat[]{{\includegraphics[width=.5\textwidth]{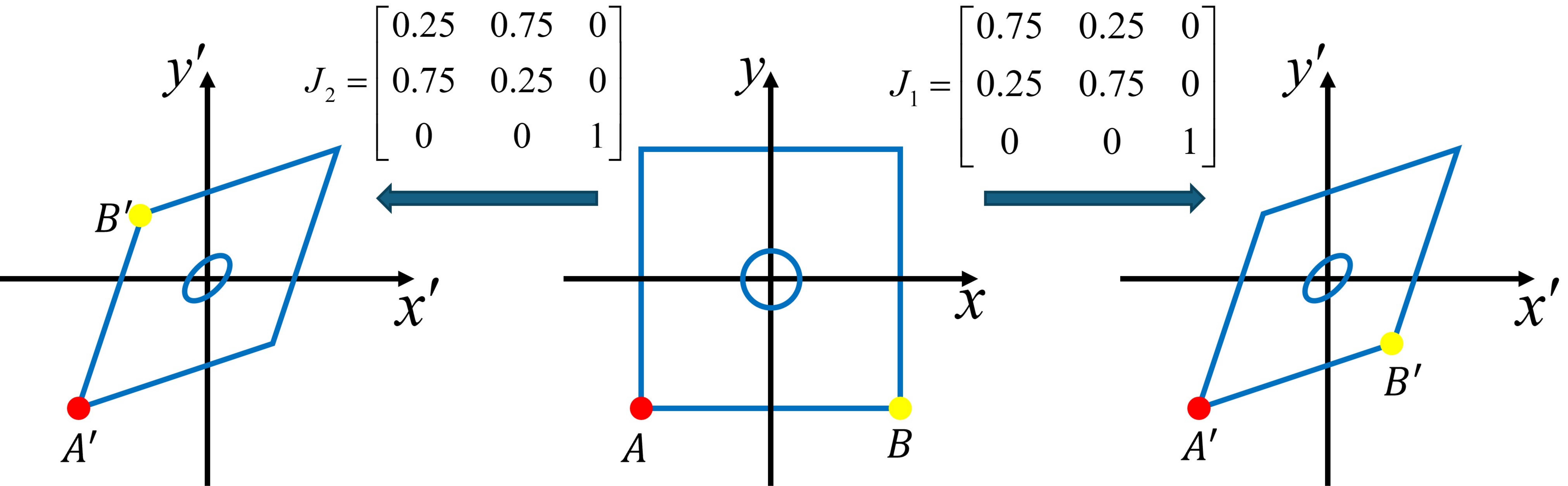}}\label{fig:figs2a}}\\

	\subfloat[]{{\includegraphics[width=.5\textwidth]{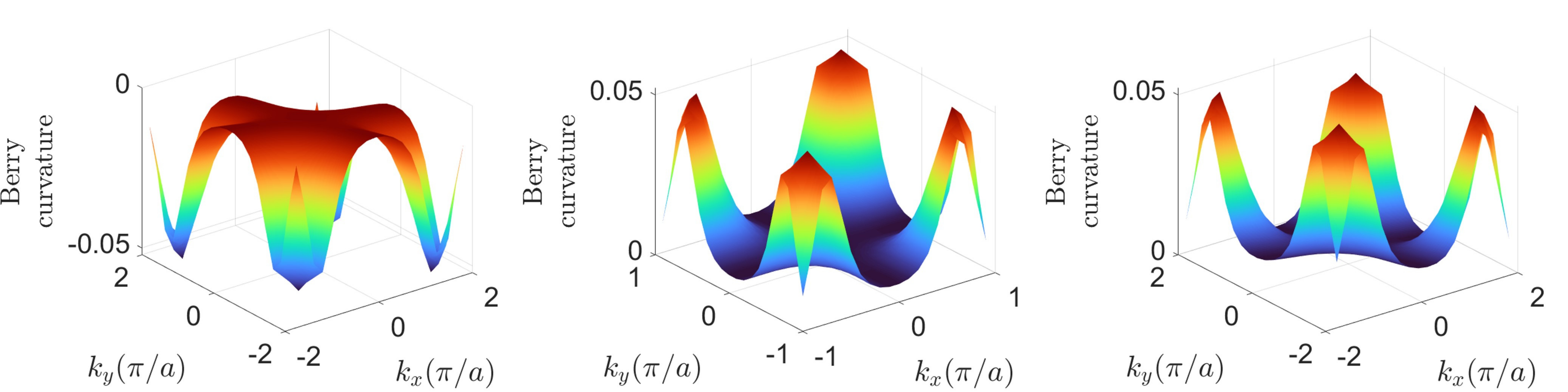}}\label{fig:figs2b}}\\
	
	\subfloat[]{{\includegraphics[width=.5\textwidth]{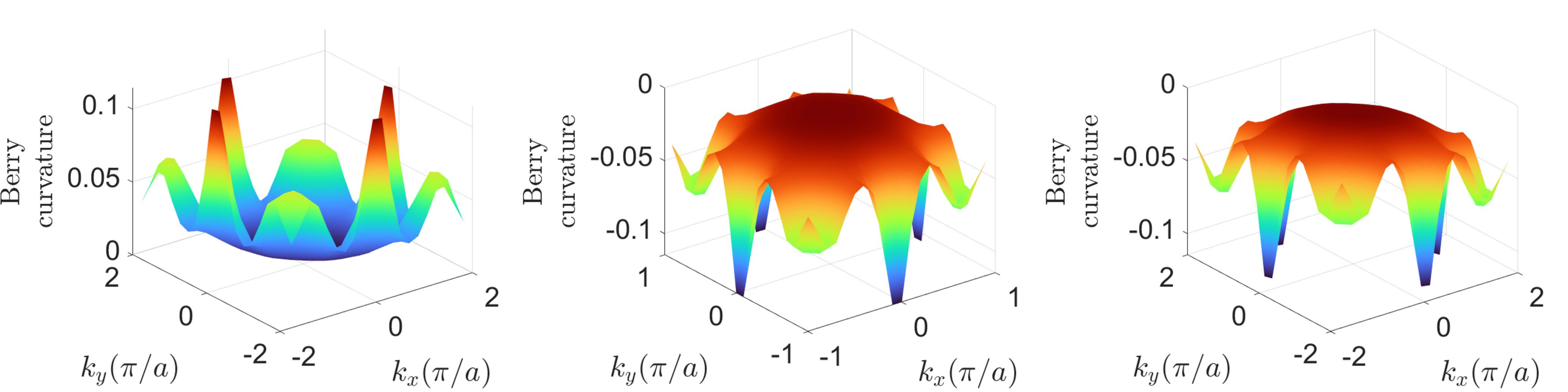}}\label{fig:figs2c}}\\

	\caption{Example of sign-changing Berry curvature. (a) Center: A square lattice with a YIG rod in the center. Right and left: The diamond lattice is transformed from the square lattice under the matrix $\bar{\bar{J}}_1$, $\bar{\bar{J}}_2$ respectively. Berry curvatures of the second band (b) and the third band (c) of corresponding structures in (a)}
\label{fig:figs2}
\end{figure}

The authors have shown the nontrivial Chern number in the gyromagnetic structure \cite{wang2008reflection}, where $C=1$ for the second band and $C=-2$ for the third band. If we apply the time-reversal transformation to the gyromagnetic material, we will change the permeability tensor $\bar{\bar{\mu}}=\begin{bmatrix}
	\mu&i\kappa&0\\
	-i\kappa&\mu&0\\
	0&0&\mu_0
\end{bmatrix}$ into $\bar{\bar{\mu}}^{\prime}=\begin{bmatrix}
	\mu&-i\kappa&0\\
	i\kappa&\mu&0\\
	0&0&\mu_0
\end{bmatrix}$ (the permittivity of the YIG rod and the background vacuum won't change). Obviously, the Chern number of the time-reversal transformed structure will change its sign since it is related to the sign of $\kappa$ in the permeability tensor. Interestingly, the transformation can also be explained from the perspective of TO. If we apply the Jacobian matrix $\bar{\bar{J}}=\begin{bmatrix}
	1&0&0\\
	0&-1&0\\
	0&0&-1
\end{bmatrix}$ to the gyromagnetic permeability tensor, we can achieve the exactly same transformed permeability tensor as the time-reversal transformation will do (again, the permittivity of the YIG rod and the background vacuum won't change). Hence, the TO shows its ability to engineer the topological invariant Chern number. 

According to Eq.~(\ref{eq:chern_transformed}), we can conclude that the Chern number in the transformed space will change its sign compared with the original space when the orientation of the Brillouin zone is flipped after the transformation. As shown in Fig. \ref{fig:figs2}, the Berry curvatures of the original structure and the transformed structures are plotted. For our linear transformation $\bar{\bar{J}}_1$, $\bar{\bar{J}}_2$, by combining Eq.~(\ref{eq:transform_k}) and Eq.~(\ref{eq:transform_chern_manu}) we can get $	C^{\prime}=\mathrm{sign}\left(\mathrm{det}\left( \dfrac{\partial\left(x^{\prime},y^{\prime} \right) }{\partial\left(x,y \right) }\right)  \right) C$. For transformation $\bar{\bar{J}}_1$, since $\mathrm{det}\left( \dfrac{\partial\left(x^{\prime},y^{\prime} \right) }{\partial\left(x,y \right) }\right) >0$, the Chern number is $C^{\prime}=1$ for the second band and $C^{\prime}=-2$ for the third band, which is the same as the original structure. However, due to $\mathrm{det}\left( \dfrac{\partial\left(x^{\prime},y^{\prime} \right) }{\partial\left(x,y \right) }\right) <0$ for $\bar{\bar{J}}_2$, the Chern number changes its sign and turns into $C^{\prime}=-1$ for the second band and $C^{\prime}=2$ for the third band. These results can be verified by observing the distributions of the Berry curvatures as shown in Fig. \ref{fig:figs2} easily.

\begin{acknowledgments}
The authors acknowledge the financial support from the National Research Foundation (grant no. NRF-CRP22-2019-0006).
\end{acknowledgments}

%
\end{document}